\documentclass[11pt,superscriptaddress,aps,prd,preprint]{revtex4-1}
\usepackage[utf8]{inputenc} 
\usepackage{amsmath}
\usepackage{mathrsfs}
\usepackage{amssymb, theorem, xspace}
\usepackage{pgfplots}
\usepackage{dcolumn} 
\usepackage{slashed}
\usepackage{appendix}
\usepackage{float}

\pgfplotsset{compat=1.13}

\newcommand{\bea}{\begin{eqnarray}}
\newcommand{\eea}{\end{eqnarray}}

\begin{document}

\title{ Heavy Quark-antiquark Systems in a Galilean Covariant MIT Bag Model}
\author{G. X. A. Petronilo}
\email{gustavopetronilo@gmail.com}

\affiliation{Universidade Federal do Pará, Campus Universitário de Salinópolis \\
Rua Raimundo Santana Cruz, São Tomé \\
68721000 - Salinópolis, PA - Brasil}

\author{M. V. T. Machado}
\email{magnus@if.ufrgs.br}
\affiliation{ Instituto de Física, Universidade Federal do Rio Grande do Sul, Porto Alegre, 91501-970, Rio Grande do Sul, Brazil}

\author{R.R. Luz}
\email{renatoluzrgs@gmail.com}
\affiliation{Departamento de Física, Universidade Estadual do Maranhão, 65054, São Luís, MA, Brazil}

\author{A.E. Santana}
\email{a.berti.santana@gmail.com}
\affiliation{International Center for Physics, Instituto de Física, Universidade de Brasília, Brasília, DF, Brazil }

\begin{abstract}
This work presents a Galilean covariant extension of the MIT bag model to describe heavy quark–antiquark systems within a non-relativistic framework. Using a 5-dimensional formulation of Galilean covariance, a Dirac-like equation governing quark dynamics inside a confining bag is constructed, and the corresponding boundary conditions are derived. Analytical solutions for the spinor wavefunctions and eigenfrequencies are obtained. Numerical results show some agreement with experimental masses and hyperfine splittings for heavy mesons, particularly charmonium and bottomonium states. However, the approach is less reliable for light-quark systems, indicating the limitations of the non-relativistic approximation.

\end{abstract}



\maketitle
\section{Introduction}

The description of hadrons as bound states of quarks and antiquarks remains one of the central problems in strong-interaction physics. Although Quantum Chromodynamics (QCD) provides the fundamental theory of the strong interaction, the nonperturbative character of confinement makes the direct analytical description of hadronic bound states a highly nontrivial task. For this reason, effective models continue to play an important role in the study of hadron spectroscopy, especially when they are able to capture the main physical features of confinement while remaining analytically and phenomenologically tractable~\cite{Gross:2022hyw,ParticleDataGroup:2024cfk,eichten1978,Ahmed:2021atj,Omugbe:2023ges,sebastian2023lienard}.

Among these effective approaches, the MIT bag model provides a simple and physically intuitive realization of confinement. In this model, quarks are assumed to move almost freely inside a finite spatial region, the bag, while suitable boundary conditions at the bag surface prevent the quark current from flowing into the exterior region. The bag constant $B$ accounts for the pressure associated with the nonperturbative QCD vacuum and contributes to the total energy of the hadronic system~\cite{chodos1974,johnson1975,arrizabalaga2019}. Despite its phenomenological nature, the bag model has been widely used to estimate hadron masses, eigenfrequencies, and hyperfine splittings.

Recent developments have extended the MIT bag framework to heavy hadrons and multiquark systems by including chromomagnetic interactions, running strong coupling effects, and enhanced binding-energy contributions. These applications indicate that bag-type models remain useful phenomenological tools for studying heavy-flavor spectroscopy and compact hadronic configurations~\cite{Zhang2021OpenHeavyBag,Yan2023HiddenHeavyTetraquarks,Zhu2024TriplyHeavyTetraquarks,Zhang2024HiddenHeavyPentaquarks}.

For systems containing heavy quarks, such as charmonium and bottomonium, a non-relativistic treatment can provide an important first approximation. The large quark masses suppress relativistic corrections in comparison with light-quark systems, making heavy quarkonia a suitable testing ground for non-relativistic effective descriptions. In this context, a formulation that preserves covariance under Galilean transformations is particularly appealing, since it allows one to describe non-relativistic dynamics in a geometrically consistent way~\cite{Takahashi,Santana,petronilo2021size,santana2001,ulhoa2009,ulhoa2011}. 

Non-relativistic and geometrical approaches, including phase-space formulations based on Wigner functions and symplectic quantum mechanics, have also been used to study quark–antiquark systems. These results suggest that alternative covariant or geometrically structured non-relativistic frameworks may provide useful complementary descriptions of heavy mesons~\cite{luz2022quark,luz2023fractional,abushady2024fractional,luz2025cornell}.

Classical physics, grounded in Galilean symmetry, serves as a fundamental framework, particularly in the study of low-energy phenomena. Despite the emergence of more general symmetries, Galilean symmetry remains a key element in the analysis of non-relativistic systems. In 1988, Takahashi and collaborators introduced a Galilean covariant formulation aimed at developing explicitly covariant non-relativistic versions of field theories. In this approach, non-relativistic dynamics is embedded in a five-dimensional manifold, allowing the Schrödinger equation to be written in a form analogous to the Klein-Gordon equation and providing a natural tensor structure for Galilean-invariant theories~\cite{Takahashi,Santana,santana2023symplectic,petronilo2019,petronilo2020,petronilo2021,Bagchi2022GalileanGauge,Baiguera2024NRQFT}.

Within this five-dimensional framework, it is also possible to construct a Dirac-like equation for non-relativistic spin-$1/2$ particles. This equation, commonly associated with the Lévy-Leblond formulation, plays the role of a non-relativistic counterpart to the Dirac equation and is directly related to the Pauli-Schrödinger dynamics~\cite{levy1967,niederer1977}. Therefore, Galilean covariance provides a suitable setting for incorporating spin degrees of freedom in non-relativistic systems while preserving a covariant structure.

The main purpose of this work is to formulate the MIT bag model within the five-dimensional Galilean covariant framework and to apply it to heavy quark–antiquark systems. Since the relative motion of heavy quarks can be treated within an approximately nonrelativistic regime, heavy quarkonia constitute a natural testing ground for improving the MIT confinement mechanism in a Galilean-covariant setting, compatible with non-relativistic sectors of QCD. Although there are different studies considering, for instance, the nonrelativistic limit of the MIT bag 
model~\cite{arrizabalaga2022,bruce2001,behrndt2024}, the Galilean covariance provides a symmetry support to explore the many aspects of the non-relativistic MIT bag formulation, as it is studied here in the application to a meson system.

Usually, the Galilean covariance is derived in a light-cone of a Minkowski space. Here, considering the effect of bag confinement, initially, we use a 6-dimensional de Sitter space, in order to demonstrate that the covariant Galilean symmetry allows us to investigate the essential features of the bag confinement consistently within the five-dimensional framework. In this construction, confined quarks are described by a Dirac-like or Lévy–Leblond-type equation, while the bag boundary condition ensures the absence of probability current through the surface. We derive the corresponding spinor solutions, obtain the eigenfrequency quantization conditions, and determine the bag constant from energy–momentum conservation. The model is then applied to heavy mesons, including a color-Coulomb interaction, similar to the Cornell potential,  and spin–spin hyperfine corrections. The results are compared with experimental masses and splittings~\cite{ParticleDataGroup:2024cfk}.

The work is organized as follows. In Section~\ref{SII}, we review the Galilean covariant formulation derived from the de Sitter construction and formulated in a five-dimensional manifold, establishing the geometrical and algebraic framework underlying the model. In Section~\ref{SIII}, we develop the Galilean covariant bag model for mesons, deriving the Dirac-like equation, the confinement boundary conditions, the spinor solutions, and the corresponding eigenfrequency quantization conditions. Section~\ref{Sect:IV} is devoted to the determination of the bag constant (B) from the energy-momentum tensor and its interpretation in terms of the pressure at the bag surface. In Section~\ref{SECV}, the model is applied to charmonium states, including the effects of the (SU(3)) gauge interaction, the color-Coulomb contribution, and spin-spin hyperfine corrections. The numerical results for heavy quarkonia and open heavy mesons are presented and discussed in Section~\ref{SVI}. Finally, Section~\ref{con} summarizes the main conclusions and discusses the applicability and limitations of the proposed Galilean covariant bag model.
\section{The De Sitter space and Galilean Covariance}\label{SII}

First, we review the formulation of Galilean covariance using an extended spacetime with five coordinates, allowing for the construction of a consistent tensor formalism. Within this framework, a Dirac-like equation (Lévy-Leblond equation) with a source with the content of a bag emerges as the non-relativistic analog of relativistic fermion dynamics. The application to the MIT Bag model is then considered in the next section.

The light cone of (4+1)-dimensional de Sitter space ($dS_5$) provides a geometric realization of the Galilei group in 4-space dimensions. This is a well-known result as in the works of Bargmann and Newton-Cartan gravity~\cite{Durval1985,Aldrovandi2003}. Here we explore this fact to introduce a field theory inside a manifold defined within a topology $S^2$, working as a bag.   

The de Sitter space $dS_5$ is a hyperboloid manifold that can be embedded in 5+1-Minkowski space, $\mathbb{R}^{5,1}$, with metric  $\eta = \mathrm{diag}(-1,1,1,1,1,1)$. In terms of components, the embedding  is such that $dX\cdot dX=\eta^{cd}dX_{c}dX_{d}=dL^2$, with $c,d=0,1,...,5$ and $X=(X_0,X_1,...,X_5)$, such that
$dL^2$ can be $dL^2>0,\,\,  dL^2<0, \,\, dL^2=0$. 
The light cone of the origin in space is the null hypersurface $\eta^{cd}dX_{c}dX_{d}=0$. Defining the new variables: $-\ell=X_5, \mathrm{u}=X_0+X_4, \mathrm{v}=X_0-X_4$ we write $d\mathrm{u}d\mathrm{v}-d\mathbf{x}^2=d\ell^2$, where $\mathbf{x}=(x^1=X_1,x^2=X_2,x^3=X_3)$. Notice that for isometry transformations, we have the $SO(4,1)$ group, and the parameter $\ell^2$ is invariant. The physical meaning of $\ell^2$ is obtained by observing that for $\mathrm{u}=0$ or $\mathrm{v}=0$, it follows that $\mathbf{x}^2=\ell^2$. Therefore, $\ell$ is the radius of a manifold $S^2$. 

Introducing the Bargmann parametrization, we write 
$
X=x=(\mathbf{x},x^4,x^5),
$
where $x^4=\mathrm{u}=\kappa_0 t, x^5=\mathrm{v}=\frac{\mathbf(x)^2}{2\kappa_0}$, where $\kappa_0 $ is a constant velocity and $t$ stands for time. Then it leads to 
\begin{equation}
d\ell^2=g_{\mu\nu}dx^{\mu}dx^{\nu},\label{3.6}
\end{equation}
with
\begin{equation}
g_{\mu\nu} = \left(
\begin{array}{ccc}
 \mathbf{1} & 0 & 0 \\
 0 & 0 & -1 \\
 0 & -1 & 0 \\
\end{array}
\right),\label{1}
\end{equation}
where $\mathbf{1}$ is a $3\times 3$ unity matrix and $\mu,\nu=1,2,...,5$. The non-null components are $g_{ij}=\delta_{ij}$, $i,j=1,2,3$, and $g_{45}=g_{54}=-1$. This is called the Galilean metric. 

The linear  coordinate transformations  are expressed as
\begin{equation}
{x^{\mu}}' = \Lambda^{\mu}{}_{\nu} x^{\nu} + a^\mu. \label{3.2}
\end{equation}
For the particular case defined by $a^\mu=({\bf a}, b,0)$ and a specific choice of $\Lambda^{\mu}{ }_\nu $, the Galilei transformations are derived. Indeed, considering   
\begin{equation}
\Lambda^{\mu}{}_\nu=
\left(\begin{array}{ccccc}
R^1_{\;1}&R^1_{\;2}&R^1_{\;3}&0&-v_1\\
R^2_{\;1}&R^2_{\;2}&R^2_{\;3}&0&-v_2\\
R^3_{\;1}&R^3_{\;2}&R^3_{\;3}&0&-v_3\\
v_iR^i_{\;1}&v_iR^i_{\;2}&v_iR^i_{\;3}&1&\frac{\textbf{v}^2}{2}\\
0&0&0&0&1
\end{array}\right),
\end{equation}
it leads to
$\overline{\mathbf{x}} = \mathcal{R}\mathbf{x+v}t+\mathbf{a,}$  and 
$\overline{t} =t+b$ , where   $\mathcal{R}=(R^i_j)$ stands for the three-dimensional Euclidean rotations, $\textbf{v}$ is the  relative velocity for Galilean boosts, $\textbf{a}$ corresponds to spatial translations, and $b$ accounts for time translations. 

In the canonical variable representation 
\begin{equation}
p_{\mu }p^{\mu } = \mathbf{p}^{2}-2mE = B^2, \label{disp5}
\end{equation}
where $p^{\mu}=(\mathbf{p},p^{4}=\kappa_0 m,p^{5}=E/\kappa_0$), such that $p_{\mu }=g_{\mu\nu}p^{\nu }$. 
In this context, Eq.~(\ref{disp5}) is the non-relativistic dispersion relation when $B=0$. 

Usually, this formalism with the Bargmann variables is explored in the light-cone, i.e., $\ell=0$ and $B=0$. Here, nevertheless, by using the De Sitter space, we have been able to provide a physical meaning for $\ell, B \neq 0$ in the full 5-dimensional manifold. Indeed, as we observed, $\ell$ is constant. It is physically interpreted by considering the  $S^2$ manifold, which, in a particular situation, is the surface of a sphere with radius $\ell$. In terms of the canonical coordinates, $B$ is also invariant and can be associated, in general, with a current in the manifold defined by the corresponding value of $\ell$. 
The fifth coordinate $ x^5 = s $ is canonically conjugated to the mass. 

Galilean physics manifests on a null-cone region in a 5-dimensional space, with $ ds = \frac {{\bf v}} {2} \cdot{\bf dr} $, consolidating the interpretations of the fifth coordinate. The Galilean transformations encompass 4-dimensional physical space embedded within a 5-dimensional framework. In this case, the basis coordinates of the vector in the 5-dimensional gradient are given by
   $ \partial_\mu\equiv \left(\nabla,\partial_t,\partial_s\right) $ such that 
    $\partial^\mu=g^{\mu\nu}\partial_\nu=(\nabla,-\partial_s,-\partial_t)\label{parial2}$. 
An invariant such as $\partial^\mu \partial_\mu =\mathcal{I}^2$
 can be used as the starting point to construct representations of a field equation accounting for a zero-spin boson field. This leads directly to the free Schrödinger equation. For a spin 1/2 fermion-field, the basic equation is $i\gamma^{\mu}\partial_{\mu}= \mathcal{I}$, where $\gamma^{\mu}$ is a representation of the Clifford algebra in 5-dimensions, i.e., $\{\gamma^{\mu},\gamma^{\nu}\}=2g^{\mu\nu}$. This leads to equations with the general form
 \begin{equation}
i\gamma^{\mu}\partial_{\mu}\psi(x)=F(\overline{\psi}(x),\psi(x)),   \label{functional2}
 \end{equation}
 where $\overline{\psi}(x)$ is the dual of $\psi(x)$  and $F(\overline{\psi}(x)\psi(x))$ is an arbitrary functional that brings information about $\ell$ and $B$.  In the next sections, we explore these geometrical results to study the confinement of quarks in mesons.  Since $\mathbf{x}$ is a 3-dimensional Euclidean space describing the configuration space, it is convenient to use the usual notation for the meaning of $\ell$ as the radius of the spatial sphere. Then we write $R=\ell$ and $r=|\mathbf{x}|=({x_1}^2+{x_2}^2+{x_3}^2)$. 

 In short, it is important to emphasize that  Galilean covariance is usually identified in the light cone of a (N+1)-Minkowski space. However, the covariant formulation  developed in this section provides a way to obtain Galilei covariance out of the light cone with a physical interpretation. This fact is explored in the sequence.

\section{The Bag Model for Galilean Mesons} \label{SIII}
In this section, we start by analyzing a meson confined in an $S^2$ space region, considering the non-relativistic sector of quantum chromodynamics, as we have discussed in the Introduction. The geometric formalism derived in the previous section is crucial here to introduce a covariant Lagrangian density describing a spin 1/2 particle. Each monomial term in the Lagrangian has a physical and mathematically well-defined structure. The results derived explicitly in this section will be explored in the remaining of the work.

\subsection{Spinor equation}
A Lagrangian density, $\mathcal{L}$, for a fermion system inside a spherical bag of radius $R$  is given by considering the linear form of the functional $F$ given in eq.~(\ref{functional2}). Then we write $\mathcal{L}$ as 
\begin{equation}
    \mathcal{L}=\bigg[\frac{1}{2}\Big(\overline{\psi}\gamma^\mu\partial_\mu\psi-(\partial_\mu\overline{\psi})\gamma^\mu\psi\Big)-B\bigg]\theta_v-\frac{1}{2}\overline{\psi}\psi\Delta_s,
\end{equation}
where  $\theta_v$ is the Heaviside step-functions,  $\theta_v=\theta(r-R)= \{1 \,\,\mathrm{for} \,\,r<R; \,\, 0\,\, \mathrm{for} \,\, r>R\}$ and $\Delta_s$ is a Dirac-delta, i.e., $\Delta_s=\delta(R-r)$, for $r<R$. It is useful to have in mind the following result 
$ \frac{\partial \theta_v}{\partial x^\mu} = \eta_\mu \Delta_s,$ 
where  $\eta_\mu$ is given by $\eta^{\mu}=(1,1,1,0,0)$ ($\eta_4=\eta_5=0$), such that  $\eta_\mu$ is a unit vector normal to the spherical surface with radius $R$.

Now we consider the 5-dimensional representation of this Clifford algebra. The matrices $\gamma^\mu$    are the Galilean Dirac matrices  written as
\begin{eqnarray}\
\gamma^i=
\left(\begin{array}{cc}
\sigma^i&0\\
0&-\sigma^i
\end{array}\right),\text{ }
\gamma^4=
\left(\begin{array}{cc}
0&0\\
\sqrt{2}&0
\end{array}\right),\text{ }
\gamma^5=
\left(\begin{array}{cc}
0&-\sqrt{2}\\
0&0
\end{array}\right).\label{matrices}
\end{eqnarray}
where $\sigma^i$ are the Pauli matrices and $\sqrt{2}$ is a short notation for a 2x2 unity matrix multiplied by the number $\sqrt{2}$.  Using these matrices, the 
dual spinor is given by  $\overline{\psi}=\zeta\psi^*$, with $\zeta=i\dfrac{\gamma_4+\gamma_5}{\sqrt{2}}$. 

To write the Euler-Lagrange equation for $ \ psi$ explicitly, we take $\mathcal{L}$   varying with $\overline{\psi}(x)$. This leads to   
\begin{equation}
    \frac{\partial \mathcal{L}}{\partial \overline{\psi}}-\frac{\partial}{\partial x^\mu}\Bigg(\frac{\partial \mathcal{L}}{\partial(\delta\overline{\psi})}\Bigg)=0,\label{E-L}
\end{equation}
where
\begin{eqnarray}
     \frac{\partial \mathcal{L}}{\partial \overline{\psi}}=\frac{1}{2}i\gamma^\mu\frac{\partial \psi}{\partial x^\mu}\theta_v-\frac{1}{2}\psi\Delta_s\label{E-L1}
\end{eqnarray}
and
\begin{eqnarray}
   \frac{\partial}{\partial x^\mu}\Bigg(\frac{\partial \mathcal{L}}{\partial(\delta\overline{\psi})}\Bigg)=  \frac{\partial}{\partial x^\mu}\Bigg[\Big(\frac{1}{2}\gamma^\mu\psi\Big)\theta_v\Bigg]=\Bigg(\frac{1}{2}\gamma^\mu\frac{\partial \psi}{\partial x^\mu}\Bigg)\theta_v-\gamma^\mu\eta_\mu\psi\Delta_s.\label{E-L2}
\end{eqnarray}
Substituting Eq.~\eqref{E-L1} and Eq.~\eqref{E-L2} into Eq.~\eqref{E-L}, the result is
\begin{eqnarray}  \Bigg(\gamma^\mu\partial_\mu\psi\Bigg)\theta_v+\frac{1}{2}\Bigg(\eta_\mu\gamma^\mu\psi-\psi\Bigg)\Delta_s=0\label{eqm}
\end{eqnarray}
Inside the bag,  $\theta_v=1$ and $\Delta_s=0$, then Eq.~\eqref{eqm} reads
\begin{equation}
    \gamma^\mu\partial_\mu\psi=0, \label{Dirac}
\end{equation}
the Galilean Dirac-Like equation. Taking Eq.~\eqref{matrices} and writing 
 $$
 \psi=\left(\begin{array}{cc}
      \varphi  \\
      \chi
 \end{array}\right), 
 $$
 Then Eq.~\eqref{Dirac} reads 
 \begin{eqnarray}
     i\sigma^i\partial_i\varphi-m\sqrt{2}\chi=0,\label{LL-1}\\
     \sigma^i\partial_i\chi-\sqrt{2}\partial_t\varphi=0\label{LL-2},
 \end{eqnarray}
 where the result $\partial_5\psi=-im\psi$ has been used. These equations are known as the Levy-Leblond equations, and they are analogous to the Dirac equation for non-Lorentzian physics. 
 
 Taking Eq.~\eqref{LL-1} and isolating $\chi$, it leads to
 \begin{equation}
     \chi=i\sigma^i\frac{\partial_i\varphi}{\sqrt{2}m}.\label{LL-3}
 \end{equation}
 Substituting the above equation into Eq.~\eqref{LL-2}, then 
 \begin{eqnarray}
    \bigg(\frac{\big(\sigma^i\partial_i\big)^2}{2m}+i\partial_t\bigg)\varphi=  \frac{\partial_i^2\varphi}{2m}+i\partial_t\varphi=0.
 \end{eqnarray}
similarly,
  \begin{eqnarray}
    \bigg(\frac{\big(\sigma^i\partial_i\big)^2}{2m}+i\partial_t\bigg)\chi=  \frac{\partial_i^2\chi}{2m}+i\partial_t\chi=0.
 \end{eqnarray}
 For the surfaces $\theta_v=0$ and $\Delta_s$, there is no-null value. Hence, it follows that 
 \begin{eqnarray}
     \frac{1}{2}\Bigg(\eta_\mu\gamma^\mu\psi-\psi\Bigg)\Delta_s=0.
 \end{eqnarray}
 As $\Delta_s\neq 0$, it leads to 
 \begin{equation}
     \eta_\mu\gamma^\mu\psi-\psi=0.\label{bc1}
 \end{equation}
From this equation, the dual is 
  \begin{equation}
     \eta_\mu\overline{\psi}\gamma^\mu+\overline{\psi}=0.\label{bc2}
 \end{equation}
 The density flow on the surface is given by
 \begin{equation}
     \eta_\mu j^\mu=\eta_\mu\overline{\psi}\gamma^\mu\psi=0.
 \end{equation}
 From Eq.~\eqref{bc1}~and ~Eq.~\eqref{bc2}, the conservation of current is derived, i.e.
 $
 \eta_\mu J^\mu=0.
 $ 
 In other words, the density flow on the surface of the bag is zero; this means the quarks exist inside the bag only. In addition, by Eq.~\eqref{matrices},  Eq.~\eqref{bc1} reads
 \begin{eqnarray}
     \eta_i\sigma^i\varphi-\varphi&=&0\\
     \eta_i\sigma^i\chi+\chi&=&0.
 \end{eqnarray} 
 At this point, it is important to analyze the boundary condition.
 \subsection{Full spinor solution and the bag boundary condition}\label{SV} 
Since only $j=1/2$ is of interest in the present work, the complete 5D light-cone bag spinor is, for $\kappa=-1$,
\begin{equation}
\psi(r,\theta,\varphi,x^4,x^5) = A \, e^{- i m x^5 - i  \frac{k^2}{2m} x^4} 
\begin{pmatrix}
ij_0(k r) \, \mathcal{Y}_{m}(\theta,\varphi) \\[1mm]
\frac{k}{\sqrt{2} m} j_{1}(k r) \, (\boldsymbol{\sigma} \cdot \hat{\mathbf{r}}) \mathcal{Y}_{ m}(\theta,\varphi)
\end{pmatrix}\label{eq.34}
\end{equation}
and for $\kappa =1$,
\begin{equation}
\psi(r,\theta,\varphi,x^4,x^5) = A \, e^{- i m x^5 - i \frac{k^2}{2m} x^4} 
\begin{pmatrix}
ij_1(k r)  \, (\boldsymbol{\sigma} \cdot \hat{\mathbf{r}}) \mathcal{Y}_{m}(\theta,\varphi) \\[1mm]
 -\frac{k}{\sqrt{2}m}j_{0}(k r) \, \mathcal{Y}_{m}(\theta,\varphi)
\end{pmatrix}. \label{eq.35}
\end{equation}

\subsection{Solution of the bag boundary condition}

Let $S=\boldsymbol{\sigma}\cdot\hat{\mathbf r}$, so that $S^2=1$, and let $Y\equiv\mathcal{Y}_m$ and $Z\equiv SY$, so that $SY=Z$ and $SZ=Y$. At $r=R$ the boundary conditions are $S\phi=\phi$ and $S\chi=-\chi$. A single-$\kappa$ solution cannot satisfy them: for $\kappa=-1$ one has $\phi\propto j_0(kr)\,Y$, and then $S\phi\propto j_0(kr)\,Z\neq\phi$. We therefore take the most general combination of the two $j=1/2$ solutions with the same $k$ (hence the same energy $k^2/2m$),
\begin{equation}
\psi = a\,\psi_{\kappa=-1} + b\,\psi_{\kappa=+1},
\end{equation}
whose upper and lower blocks, with $c=k/(\sqrt{2}\,m)$, are
\begin{eqnarray}
\phi &=& i\big[\,a\,j_0(kr)\,Y + b\,j_1(kr)\,Z\,\big],\\
\chi &=& c\,\big[\,a\,j_1(kr)\,Z - b\,j_0(kr)\,Y\,\big].
\end{eqnarray}

Imposing the conditions at $r=R$ and using $SY=Z$, $SZ=Y$, we find
\begin{eqnarray}
S\phi=\phi &\;\Rightarrow\;& a\,j_0(kR)-b\,j_1(kR)=0,\\
S\chi=-\chi &\;\Rightarrow\;& a\,j_1(kR)-b\,j_0(kR)=0.
\end{eqnarray}
This is a homogeneous linear system for the mixing coefficients $(a,b)$:
\begin{equation}
\begin{pmatrix} j_0(kR) & -j_1(kR)\\ j_1(kR) & -j_0(kR)\end{pmatrix}
\begin{pmatrix} a\\ b\end{pmatrix}=0 .
\end{equation}
A nontrivial solution exists only if the determinant vanishes,
\begin{equation}
-j_0^2(kR)+j_1^2(kR)=0
\;\;\Longrightarrow\;\;
j_0(kR)=\pm\, j_1(kR).
\label{eq:j0j1}
\end{equation}
Eq.~\eqref{eq:j0j1} is therefore not an additional assumption: it is the condition for the boundary conditions to admit a nonzero spinor. The sign is tied to the ratio of the coefficients, since the first equation gives $b/a=j_0(kR)/j_1(kR)=\pm1$.

\paragraph{Eigenvalue equations.} With $x=kR$, $j_0(x)=\sin x/x$ and $j_1(x)=\sin x/x^2-\cos x/x$, the two signs in Eq.~\eqref{eq:j0j1} give
\begin{center}
\begin{tabular}{llll}
\hline
Condition & Mixing & Eigenvalue equation & Lowest roots \\
\hline
$j_0=+j_1$ & $a=b$  & $\tan x = \dfrac{x}{1-x}$ & $x\simeq 2.04,\;5.40,\;8.58,\dots$ \\[2mm]
$j_0=-j_1$ & $a=-b$ & $\tan x = \dfrac{x}{1+x}$ & $x\simeq 3.81,\;7.0,\dots$ \\
\hline
\end{tabular}
\end{center}
Both signs are consistent solutions of the boundary condition. The $+$ branch contains the lowest eigenvalue, $x\simeq 2.04$, which is the ground state; the $-$ branch gives a separate tower of higher states.

\paragraph{Ground-state spinor.} For $a=b=N$ and $k=2.04/R$,
\begin{equation}
\psi = N\,e^{-imx^5-i\frac{k^2}{2m}x^4}
\begin{pmatrix}
i\,\big[j_0(kr)+j_1(kr)\,\boldsymbol{\sigma}\!\cdot\!\hat{\mathbf r}\big]\mathcal{Y}_m\\[1mm]
\dfrac{k}{\sqrt{2}\,m}\,\big[j_1(kr)\,\boldsymbol{\sigma}\!\cdot\!\hat{\mathbf r}-j_0(kr)\big]\mathcal{Y}_m
\end{pmatrix}.
\end{equation}
At $r=R$ one has $j_0=j_1\equiv j$, so $\phi=ij(1+S)Y$ and $\chi=cj(S-1)Y$. Then $S\phi=ij(S+1)Y=\phi$ and $S\chi=cj(1-S)Y=-\chi$, so both conditions are satisfied. The energy is
\begin{equation}
E=\frac{k^2}{2m}=\frac{x^2}{2mR^2}\simeq\frac{2.08}{mR^2}.
\end{equation}

In the following, the meaning and the value of the constant $B$ are addressed.
\section{The Value of The Constant B}\label{Sect:IV}
The value of the bag constant $B$ is determined by considering the energy-momentum tensor and using its symmetry properties. The Energy-momentum tensor is given by
\begin{equation}
    T^{\mu\nu}=\Bigg[\left(\partial^\nu\overline{\psi}\right)\frac{\partial \mathcal{L}}{\partial(\partial_\mu\overline{\psi})}+\left(\partial^\nu{\psi}\right)\frac{\partial \mathcal{L}}{\partial(\partial_\mu{\psi})}\Bigg]-g^{\mu\nu}\mathcal{L}.\label{E-M}
\end{equation}
The two terms inside the brackets lead to
\begin{equation}
    \Bigg[\left(\partial^\nu\overline{\psi}\right)\frac{\partial \mathcal{L}}{\partial(\partial_\mu\overline{\psi})}+\left(\partial^\nu{\psi}\right)\frac{\partial \mathcal{L}}{\partial(\partial_\mu{\psi})}\Bigg]=\frac{i}{2}\Big(\overline{\psi}\gamma^\mu\partial^\nu\psi-(\partial^\nu\overline{\psi})\gamma^\mu\psi\Big)\theta_v,\label{term1}
\end{equation}
The last term is given by
\begin{equation}
    -g^{\mu\nu}\mathcal{L}=-g^{\mu\nu}\Bigg(\frac{i}{2}\Big(\overline{\psi}\gamma^\lambda\partial_\lambda\psi-(\partial_\lambda\overline{\psi})\gamma^\lambda\psi\Big)\theta_v+\Big(B\theta_v-\frac{1}{2}\overline{\psi}\psi\Delta_s\Big)\Bigg).\label{term2}
\end{equation}
Substituting Eq.~\eqref{term1} and Eq~\eqref{term2} into Eq.~\eqref{E-M}, it follows
\begin{align}
    T^{\mu\nu}&=\frac{i}{2}\Big(\overline{\psi}\gamma^\mu\partial^\nu\psi-(\partial^\nu\overline{\psi})\gamma^\mu\psi\Big)\theta_v-g^{\mu\nu} \nonumber
    \\&\times\Bigg(\Big(\frac{i}{2}\Big(\overline{\psi}\gamma^\lambda\partial_\lambda\psi-(\partial_\lambda\overline{\psi})\gamma^\lambda\psi-B\Big)\theta_v-\frac{1}{2}\overline{\psi}\psi\Delta_s\Bigg).
\end{align}
By conservation, we have
\begin{eqnarray}
    \partial_\mu T^{\mu\nu}&=&\frac{i}{2}\Big(\overline{\psi}\gamma^\mu\partial_\mu\partial^\nu\psi-(\partial_\mu\partial^\nu\overline{\psi})\gamma^\mu\psi\Big)\theta_v \nonumber\\
    &&-\frac{i}{2}\Big(\overline{\psi}\gamma^\lambda\partial^\nu\partial_\lambda\psi-(\partial^\nu\partial_\lambda\overline{\psi})\gamma^\lambda\psi\Big)\theta_v+\partial_\mu\Big(g^{\mu\nu}B\theta_v+\frac{1}{2}g^{\mu\nu}\overline{\psi}\psi\Delta_s\Big),\nonumber
\end{eqnarray}
where we used $g^{\mu\nu}\partial_\mu=\partial^\nu$ and the fact that $\gamma^\mu\partial_\mu$ and $\gamma^\lambda\partial_\lambda$ are summed over; thus, we can subtract both 
\begin{eqnarray}
    \partial_\mu T^{\mu\nu} &=& \partial_\mu\Big(g^{\mu\nu}B\theta_v+\frac{1}{2}g^{\mu\nu}\overline{\psi}\psi\Delta_s\Big)\nonumber\\
    &=&g^{\mu\nu}\Big(\eta_\mu B\Delta_s+\frac{1}{2}\partial_\mu\big(\overline{\psi}\psi\big)\Delta_s\Big)\nonumber\\
    &=&\eta^\nu B\Delta_s+\frac{1}{2}\partial^\nu\big(\overline{\psi}\psi\big)\Delta_s=0,
\end{eqnarray}
where $\partial_\mu \Delta_s=0$ and we use the raising and lowering of indices in the last line. As the quantity $\Delta_s$ is only nonzero on the surface, we have
\begin{equation}
    \eta^\nu B=-\frac{1}{2}\partial^\nu\big(\overline{\psi}\psi\big).
\end{equation}
Using the fact that $\eta^\nu\eta_\nu=\boldsymbol{\eta}^2=1$, we obtain
\begin{equation}
    B=-\frac{1}{2}\eta_\nu\partial^\nu\big(\overline{\psi}\psi\big)
\end{equation}
Since the bag has been taken as spherically symmetric, it follows that
\begin{equation}
    B=-\frac{1}{2}\frac{\partial}{\partial r}\left(\overline{\psi}\psi\right),
\end{equation}
Therefore, we interpret the constant B as the pressure on the surface of the bag. 

Using the Energy-Momentum tensor, we can construct the energy of the Bag. The field momentum density in Galilean Covariance is written as~\cite{santos2004galilean} 
\begin{equation}
    P^\mu=\int d^3x T^{4\mu}.
\end{equation}
Therefore, the energy of the Bag is given by
\begin{eqnarray}
        P^5&=&-\int d^3x T^{45}\nonumber\\
        &=& \int d^3x\, \Bigg(\frac{i}{2}\Big(\overline{\psi}\gamma^4\partial_4\psi-(\partial_4\overline{\psi})\gamma^4\psi\Big)\theta_v -\frac{i}{2}\Big(\overline{\psi}\gamma^\mu\partial_\mu\psi-(\partial_\mu\overline{\psi})\gamma^\mu\psi\Big)\theta_v+(B\theta_v +\frac{1}{2}(\overline{\psi}\psi)\Delta_s)\Bigg).\nonumber
\end{eqnarray}
Hence, inside the Bag, we have
\begin{eqnarray}
    P^5=E=\int d^3x\,\bigg( -\frac{i}{2}\Big(\overline{\psi}\gamma^i\partial_i\psi-(\partial_i\overline{\psi})\gamma^i\psi\Big)-\bigg(\frac{i}{2}\Big(\overline{\psi}\gamma^5\partial_5\psi-(\partial_5\overline{\psi})\gamma^5\psi\Big)-B\bigg)
\end{eqnarray}

and the mass density is
\begin{eqnarray}
    P^4=M=\int d^3x\,\bigg( -\frac{i}{2}\Big(\overline{\psi}\gamma^i\partial_i\psi-(\partial_i\overline{\psi})\gamma^i\psi\Big)-\bigg(\frac{i}{2}\Big(\overline{\psi}\gamma^4\partial_4\psi-(\partial_4\overline{\psi})\gamma^4\psi\Big)-B\bigg)
\end{eqnarray}

In the next section, these results are used to analyze a meson state
\section{Charmonium states} \label{SECV}

In this section, we estimate the energies of heavy quark--antiquark $S$-wave
states in the Galilean covariant bag. We first derive the energy and the
equilibrium condition that fixes the bag radius, then add the gluon-exchange
effects through minimal $SU(3)_c$ coupling and finally fix the model
parameters and present the charmonium results. Throughout, $m_c=1.3$~GeV and
$m_b=4.9$~GeV.

\subsection{Energy and equilibrium radius}

In the Galilean framework, the mass is a central extension of the algebra, so
the total mass $M=m_{q_1}+m_{q_2}$ is conserved, and the center-of-mass motion
decouples exactly from the internal motion. With
\begin{equation}
\mathbf R_{\rm cm}=\frac{m_{q_1}\mathbf r_1+m_{q_2}\mathbf r_2}{M},\qquad
\mathbf r=\mathbf r_1-\mathbf r_2,\qquad
\frac1\mu=\frac1{m_{q_1}}+\frac1{m_{q_2}},
\end{equation}
The kinetic term splits as 
\begin{equation}
\frac{\mathbf p_1^2}{2m_{q_1}}+\frac{\mathbf p_2^2}{2m_{q_2}}
=\frac{\mathbf P^2}{2M}+\frac{\mathbf p^2}{2\mu}.
\end{equation}
In the center-of-mass frame ($\mathbf P=0$), the energy is purely internal.
Both constituents occupy the lowest bag orbital of Sec.~\ref{SV}, with
$k=\omega/R$, where $\omega=kR\simeq2.043$ is the lowest root of
$j_0(\omega)=j_1(\omega)$. The kinetic energy is then
\begin{equation}
\frac{\omega^2}{2R^2}\Big(\frac1{m_{q_1}}+\frac1{m_{q_2}}\Big)=\frac{\omega^2}{2\mu R^2},
\end{equation}
and the energy of the bag is
\begin{equation}
E_0(R)=m_{q_1}+m_{q_2}+\frac{\omega^2}{2\mu R^2}+\frac{4\pi}{3}BR^3-\frac{Z_{\rm eff}}{R},
\qquad
Z_{\rm eff}\equiv Z_0+\frac43\,\alpha_s .
\label{eq:E0}
\end{equation}
The terms are, in order, the rest masses, the kinetic energy, the volume
energy, and an attractive contribution consisting of a phenomenological zero-point
term $Z_0$ and the color-Coulomb term $(4/3)\alpha_s$ derived in
Sec.~\ref{SecCoul}. Both have the same $1/R$ dependence and are therefore
combined into $Z_{\rm eff}$. The condition $dE_0/dR=0$ gives the pressure balance
\begin{equation}
-\frac{\omega^2}{\mu R^3}+4\pi BR^2+\frac{Z_{\rm eff}}{R^2}=0
\;\;\Longleftrightarrow\;\;
4\pi BR^5+Z_{\rm eff}R=\frac{\omega^2}{\mu}.
\label{eq:balance}
\end{equation}
The left-hand side increases monotonically with $R$, so the equilibrium radius
is unique for given $B$, $Z_{\rm eff}$, and masses. If $\alpha_s$ depends on $R$,
Eq.~\eqref{eq:balance} is solved self-consistently. For a bag with nonzero total
momentum, the energy is $E_0(R)+\mathbf P^2/2M$, and the equilibrium radius is
unchanged.

\subsection{Minimal Gauge Coupling and Hyperfine Splitting}

Gluon effects are introduced through the minimal $SU(3)_c$ coupling,
\begin{equation}
\partial_\mu\rightarrow D_\mu=\partial_\mu-igA^a_\mu T^a ,
\end{equation}
where $g$ is the color coupling, $A^a_\mu$ is the gluon field, and $T^a$
($a=1,\dots,8$) are the $SU(3)_c$ generators. For the spatial part,
\begin{equation}
(\boldsymbol\sigma\!\cdot\!\mathbf D)^2=\mathbf D^2+i\,\boldsymbol\sigma\!\cdot\!(\mathbf D\times\mathbf D)
=\mathbf D^2+g\,\boldsymbol\sigma\!\cdot\!\mathbf B^aT^a ,
\end{equation}
since $\mathbf D\times\mathbf D=-ig\,\mathbf B^aT^a$. The non-commutativity of
the covariant derivatives therefore generates a chromo-magnetic (Pauli)
interaction of strength $g/2m$ in the Hamiltonian. Its exchange between the
quark and the antiquark gives, in Coulomb gauge (Appendix~A), the effective
spin--spin interaction
\begin{equation}
H_{SS}=\frac{32\pi\alpha_s}{9\,m_{1}m_{2}}\;
\boldsymbol S_1\!\cdot\!\boldsymbol S_2\;\delta^{(3)}(\mathbf r_1-\mathbf r_2),
\qquad \alpha_s=\frac{g^2}{4\pi}.
\label{eq:HSS}
\end{equation}
This term is treated to first order and does not modify the equilibrium
condition~\eqref{eq:balance}.

\paragraph{Contact density.}
Both constituents occupy the same orbital, so the expectation value of the
contact operator is
\begin{equation}
\rho_c\equiv\big\langle\delta^{(3)}(\mathbf r_1-\mathbf r_2)\big\rangle
=\int d^3r\,|\phi(\mathbf r)|^4,\qquad \int d^3r\,|\phi|^2=1,
\label{eq:rhoc}
\end{equation}
where $\phi$ is the upper component of the spinor, which carries the
probability density of the L\'evy-Leblond equation. For the ground state of
Sec.~\ref{SV}, $\phi=iN\,[\,j_0(kr)+j_1(kr)\,\boldsymbol\sigma\!\cdot\!\hat{\mathbf r}\,]\,\mathcal Y_m$
with $\mathcal Y_m=\chi_m/\sqrt{4\pi}$. Using
$\chi^\dagger_\pm(\boldsymbol\sigma\!\cdot\!\hat{\mathbf r})\chi_\pm=\pm\cos\theta$,
\begin{equation}
|\phi|^2=\frac{N^2}{4\pi}\Big[A(r)\pm B(r)\cos\theta\Big],\qquad
A=j_0^2+j_1^2,\quad B=2j_0j_1 .
\end{equation}
The cross term integrates to zero over angles, so $N^2=k^3/I_2$ with
\begin{equation}
I_2=\int_0^{\omega}\big(j_0^2+j_1^2\big)u^2\,du\simeq1.654 .
\end{equation}
Using $\int_{-1}^{1}(A\pm B\cos\theta)^2\,d\cos\theta=2A^2+\tfrac23B^2$,
\begin{equation}
\rho_c=\frac{\omega^3}{4\pi R^3}\frac{I_4}{I_2^{\,2}},\qquad
I_4=\int_0^{\omega}\Big[\big(j_0^2+j_1^2\big)^2+\tfrac43\,j_0^2j_1^2\Big]u^2\,du\simeq1.25,
\end{equation}
that is,
\begin{equation}
\rho_c\simeq\frac{0.309}{R^3}=f\,\frac{3}{4\pi R^3},\qquad f\simeq1.29 .
\label{eq:rhocnum}
\end{equation}
The contact density is $29\%$ larger than that of a uniform sphere, and the
factor $f$ is independent of $R$ and of the quark masses.

\paragraph{Hyperfine energy.}
For $S$-wave states $\langle\boldsymbol S_1\!\cdot\!\boldsymbol S_2\rangle=+1/4$
($S=1$) and $-3/4$ ($S=0$), so
\begin{equation}
E(S)=E_0+\Delta E_{\rm HF}\,\langle\boldsymbol S_1\!\cdot\!\boldsymbol S_2\rangle,
\qquad
\Delta E_{\rm HF}=\frac{32\pi\alpha_s}{9\,m_1m_2}\,\rho_c\simeq\frac{3.45\,\alpha_s}{m_1m_2R^3}.
\label{eq:DEHF}
\end{equation}
The $S=1$--$S=0$ mass splitting is exactly $\Delta E_{\rm HF}$.

\subsection{Coulomb-like Interaction and Running Coupling}\label{SecCoul}

Minimal coupling also generates the color-Coulomb potential from one-gluon
exchange,
\begin{equation}
V_C(r)=-\frac43\frac{\alpha_s}{r}.
\end{equation}
For the ground state we use $\langle 1/r\rangle\simeq 1/R$, so that
\begin{equation}
\Delta E_C(R)=-\frac43\frac{\alpha_s}{R}.
\end{equation}
This term has the same $R$-dependence as the zero-point term $-Z_0/R$ and is
therefore part of the total energy~\eqref{eq:E0} and of the equilibrium
condition~\eqref{eq:balance} through $Z_{\rm eff}=Z_0+\tfrac43\alpha_s$. It is not
added afterwards as a perturbative correction: for $\alpha_s\sim0.2$--$0.4$ it
changes the equilibrium radius by about $10$--$15\%$. The approximation
$\langle1/r\rangle\simeq1/R$ is crude and affects the Coulomb coefficient at the
level of tens of percent.

Asymptotic freedom implies that $\alpha_s$ depends on the interquark scale.
At leading order,
\begin{equation}
\alpha_s(Q^2)=\frac{4\pi}{\beta_0\ln(Q^2/\Lambda_{\rm QCD}^2)},\qquad
\beta_0=11-\frac23N_f ,
\label{eq:alphas}
\end{equation}
with $\Lambda_{\rm QCD}=220$~MeV and $N_f=3$ for all systems~\cite{Gross:2022hyw}.
For simplicity, $\alpha_s$ is evaluated once for each meson at the scale
\begin{equation}
Q=\rho_Q\,(m_{q_1}+m_{q_2}),
\label{eq:Qrule}
\end{equation}
where $\rho_Q$ is a universal parameter.
The one-loop expression is not reliable for $Q$ of a few $\Lambda_{\rm QCD}$,
which limits the model to heavy quarks.
\subsection{Model parameters}
The model has three universal parameters: the bag constant $B$, the
zero-point coefficient $Z_0$, and the scale parameter $\rho_Q$. They are fixed
once for all systems by a joint least-squares fit to the masses and hyperfine
splittings of $J/\psi$, $\eta_c$, $\Upsilon(1S)$, and
$\eta_b$~\cite{ParticleDataGroup:2024cfk}, with a $1\%$ uncertainty for the
spin-averaged masses and $10$~MeV for the splittings. The fit gives
$
B^{1/4}=0.343~\mathrm{GeV}\;(B\simeq1.81~\mathrm{GeV/fm^3}),\;
Z_0=1.08,\; \rho_Q=0.61 ,
$
which corresponds to $\alpha_s=0.35$ for $c\bar c$ and $0.21$ for $b\bar b$. All other mesons are
predictions. The zero-point coefficient $Z_0=1.08$ is lower than the
light-hadron MIT value $1.84$~\cite{DeGrand:1975cf}. With $Z_0=1.84$ held fixed,
no universal choice of $B$ and $\rho_Q$ reproduces both charmonium and
bottomonium: the fit drives $\alpha_s\to0$ and leaves a $4.5\%$ rms error in the masses.
The only quantities that are not fitted are the masses of the other mesons.

The bag radius $R\simeq0.36$~fm is of the order of the rms $c\bar c$ separation
of about $0.4$--$0.5$~fm obtained in potential models and is larger than
the charge radius~\cite{Akbar:2023eqm}, as expected for a confinement radius.
It is the location of the bag surface, not an rms or charge radius, and
should not be compared directly with them. We emphasize three limitations.

First, the fitted bag constant $B^{1/4}=343$~MeV is about $2.4$ times the typical
value $B^{1/4}\simeq145$~MeV of baryon bag fits~\cite{Haxton:1980mc,Hasenfratz:1980jv}.
The model therefore does not support a $B$ that is independent of the hadron
class. Second, $Z_0$ and the scale parameter $\rho_Q$ are also fitted, so the
agreement for charmonium and bottomonium is not a prediction. Only the
other mesons of Sec.~\ref{SVI} provide tests. Third, the ground-state spinor of
Sec.~\ref{SV} mixes the $\kappa=\mp1$ solutions of opposite parity, which
produces an angular dependence of the density in Eq.~\eqref{eq:rhoc}.
The contact factor $f=1.29$ depends on this choice.

\section{Results and Discussion} \label{SVI}

\begin{table}[H]
\centering
\caption{Results for all systems with the universal parameters $B^{1/4}\approx 343$~MeV, $Z_0\approx 1.08$, $\rho_Q\approx 0.61$, $m_c=1.3$~GeV, $m_b=4.9$~GeV, $m_s=0.095$~GeV. The coupling scale is $Q=\rho_Q(m_{q_1}+m_{q_2})$, with a lower cutoff $Q\geq2$~GeV for the systems with a strange quark. The $c\bar c$ and $b\bar b$ rows were fitted; the $b\bar c$ row is a prediction; rows marked $^\dagger$ are outside the domain of the model. Experimental splittings are computed from the PDG masses; the $B_c^*$ mass is not measured.}
\vskip 8pt
\label{tab:all}
\resizebox{\textwidth}{!}{
\begin{tabular}{|l|c|c|c|c|c|c|c|c|c|}
\hline
System & $Q$ [GeV] & $\alpha_s$ & $R$ [fm] & $E(S{=}1)$ [GeV] & $E(S{=}0)$ [GeV] & $\Delta E_{\rm HF}$ [MeV] & $\Delta E_{\rm HF}^{\rm ref}$ [MeV] & Err$(S{=}1)$ [\%] & Err$(S{=}0)$ [\%] \\
\hline
$c\bar c$ ($J/\psi,\eta_c$) & 1.59 & 0.353 & 0.361 & 3.100 & 2.982 & 117.5 & 113.0 & 0.1 & 0.1 \\
$b\bar b$ ($\Upsilon,\eta_b$) & 5.99 & 0.211 & 0.212 & 9.352 & 9.328 & 24.6 & 61.6 & 1.1 & 0.8 \\
$b\bar c$ ($B_c^*,B_c$) & 3.79 & 0.245 & 0.315 & 6.363 & 6.330 & 32.6 & 55.5 & 0.5 & 0.9 \\
\hline
$s\bar c$ ($D_s^*,D_s$)$^\dagger$ & 2.00 & 0.316 & 0.592 & 5.169 & 4.842 & 326.4 & 143.7 & 144.7 & 146.0 \\
$s\bar b$ ($B_s^*,B_s$)$^\dagger$ & 3.05 & 0.265 & 0.586 & 8.596 & 8.521 & 74.9 & 48.9 & 58.7 & 58.8 \\
$s\bar s$ ($\phi,\eta'$)$^\dagger$ & 2.00 & 0.316 & 0.676 & 6.588 & 3.587 & 3000.8 & 61.7 & 546.2 & 274.5 \\
\hline
\end{tabular}}
\end{table}

 Some discussion on the results presented in Tab. \ref{tab:all} is in order. First, a common effective parameter set can organize the gross heavy-sector mass scale and the expected radius ordering, and it describes the calibrated charmonium doublet well. The substantial bottomonium hyperfine deficit, the sensitivity to the boundary wavefunction, and the marginal charm expansion parameter prevent a broader claim of quantitatively controlled heavy-quark spectroscopy. The mixed-flavor result is useful as an independent test. The Galilean formulation supplies a geometrical and spinorial organization of the nonrelativistic bulk theory. The spectral energies used here, nevertheless, reduce to a cavity kinetic term supplemented by phenomenological volume, zero-point, Coulomb, and contact interactions. 

The mass centroids and hyperfine splittings can be discussed in more detail.  For a common spin-independent radius, the leading contact interaction gives
\begin{align}
 M_V&=E_0+\frac14\Delta_{\mathrm{HF}},&
 M_P&=E_0-\frac34\Delta_{\mathrm{HF}},\\
 \overline M_{1S}&\equiv\frac{3M_V+M_P}{4}=E_0,&
 M_V-M_P&=\Delta_{\mathrm{HF}}.
 \label{eq:rd_centroid}
\end{align}
The centroid removes the first-order spin--spin contribution and isolates the confinement balance. The splitting then tests the contact matrix element separately. Table~\ref{tab:rd_heavy} reorganizes the present results in this form. Centroids are calculated from the rounded masses in the present work; splittings retain the separately tabulated values and need not equal the difference of the displayed masses exactly.

\begin{table}[t]
\centering
\caption{Heavy-sector results of the present implementation. The $c\bar c$ and $b\bar b$ systems enter the fit; $b\bar c$ is a prediction. The reference $b\bar c$ splitting is theoretical. The final column is the signed relative splitting residual, $100(\Delta_{\mathrm{HF}}/\Delta_{\mathrm{ref}}-1)$.}
\label{tab:rd_heavy}
\small
\begin{tabular}{|lrrrrr|}
\toprule
System & $R$ [fm] & $\overline M_{1S}$ [GeV] & $\Delta_{\mathrm{HF}}$ [MeV] & $\Delta_{\mathrm{ref}}$ [MeV] & Residual [\%]\\
\hline
$c\bar c$ & 0.361 & 3.0705 & 117.5 & 113.0 & $+4.0$\\
$b\bar b$ & 0.212 & 9.3460 & 24.6 & 61.6 & $-60.1$\\
$b\bar c$ & 0.315 & 6.3548 & 32.6 & 55.5 & $-41.3$\\
\hline
\end{tabular}
\end{table}

In charmonium, the predicted vector and pseudoscalar masses, $3.100$ and $2.982$ GeV, reproduce the reference masses at the quoted $0.1\%$ level. The splitting exceeds the reference by only $4.5$ MeV. Within the stated $10$ MeV fitting scale, this satisfactorily describes the calibrated charmonium doublet. Nevertheless, the result remains conditional on the adopted mode, contact factor, and quark-mass convention. Bottomonium provides a more restrictive test. The masses $9.352$ and $9.328$ GeV lie below the reference values, with the quoted errors of $1.1\%$ and $0.8\%$, respectively. More decisively, the splitting is only $24.6$ MeV, compared with $61.6$ MeV in the reference comparison. The $37.0$ MeV deficit is $3.7$ times the adopted $10$ MeV fitting scale and corresponds to a $60\%$ relative underestimate. Thus, the universal parameter set does not reproduce the spin dependence of both heavy-quarkonium families simultaneously. Describing the fit solely through the relative errors in the total masses would conceal this failure.

The predicted $b\bar c$ masses are $M_{B_c^*}=6.363$ GeV and $M_{B_c}=6.330$ GeV. The pseudoscalar result differs from the measured reference by about $0.9\%$, as shown in Tab. \ref{tab:all}. The predicted splitting of $32.6$ MeV is smaller than the adopted theoretical benchmark by $22.9$ MeV. This is a tension with that benchmark, not a discrepancy with an observed $B_c^*$ mass. The mixed-flavor sector is, therefore, promising as an independent constraint.

Concerning the obtained equilibrium radii, they satisfy the condition
\begin{equation}
 R_{b\bar b}<R_{b\bar c}<R_{c\bar c},
 \qquad (0.212,\,0.315,\,0.361)~\mathrm{fm}.
 \label{eq:rd_radii}
\end{equation}
This ordering follows naturally from the reduced-mass dependence of the kinetic energy. Actually, if the attractive term $-Z_{\mathrm{eff}}/R$ were absent, the radius would scale as $R\propto(B\mu)^{-1/5}$. The attraction reduces the equilibrium radius, and its relative importance changes with flavor. This makes the decreasing radius with increasing reduced mass physically plausible. However, the cavity radius is a boundary parameter, not an RMS interquark separation or a charge radius.

As noted before, the fitted bag scale is much larger than the conventional $B^{1/4}\simeq145$ MeV value quoted for light-hadron fits. Their ratio is about $2.37$ at the level of $B^{1/4}$ and about $31$ at the level of the energy density $B$. The fitted value corresponds to approximately $1.80~\mathrm{GeV/fm^3}$. This comparison does not rule out an effective heavy-sector parametrization because the surface prescription and other terms differ between models. It does mean that the present fit does not demonstrate a universal vacuum energy density across light and heavy hadrons.

Now, let us discuss the flavor dependence of the contact interaction. The present contact prescription gives
\begin{equation}
 \Delta_{\mathrm{HF}}=
 \frac{32\pi\alpha_s}{9m_1m_2}\rho_c,
 \qquad \rho_c\simeq\frac{0.309}{R^3},
 \qquad \Delta_{\mathrm{HF}}\simeq
 \frac{3.45\alpha_s}{m_1m_2R^3},
 \label{eq:rd_hyperfine}
\end{equation}
with $R$ expressed in $\mathrm{GeV}^{-1}$ in the final numerical expression. Three effects compete: the explicit inverse-mass suppression, the decrease of $\alpha_s$ with the chosen scale, and the enhancement of the contact density as the bag shrinks. The smaller bottomonium radius partially compensates for the $1/m_b^2$ factor, but not enough to reproduce its observed splitting. For equal-flavor systems, the ratio is
\begin{equation}
 \frac{\Delta_{b\bar b}}{\Delta_{c\bar c}}
 =\frac{\alpha_s^{(b)}}{\alpha_s^{(c)}}
 \left(\frac{m_c}{m_b}\right)^2
 \left(\frac{R_{c\bar c}}{R_{b\bar b}}\right)^3
 \simeq0.21,
 \label{eq:rd_splittingratio}
\end{equation}
whereas the reference ratio is $61.6/113.0\simeq0.55$. Since the two fitted systems share the same dimensionless contact coefficient, this discrepancy exposes a limitation of their assumed flavor dependence. It cannot be removed by a common multiplicative enhancement of the contact term while preserving the charmonium agreement. At fixed radius and coupling, bottomonium would require a contact enhancement of about $61.6/24.6\simeq2.50$. Alternatively, changing the radius alone would require $R_{b\bar b}\simeq0.156$ fm, about $26\%$ below its present value. 

However, there is a further qualification. The upper component used here combines the $\kappa=-1$ and $\kappa=+1$ modes, producing an angular dependence in its density. The coefficient $0.309$ follows from that specific construction. For a general spinor orbital, the two-body matrix element must be evaluated in states coupled to definite total angular momentum and parity. Replacing it with a scalar overlap times $\langle\mathbf S_1\cdot\mathbf S_2\rangle$ assumes that the spatial and spin factors separate appropriately. The present numerical results should therefore be read as conditional on this factorization. A direct two-body calculation would test whether the same contact coefficient applies to both members of the doublet and whether additional spin-angular terms survive.

Finally, we comment on the results for strange-quark systems as extrapolation tests. The rows containing strange quarks in Tab. \ref{tab:all} display the breakdown of extrapolating the same nonrelativistic treatment to a current-like mass $m_s = 0.095$ GeV. The predicted $D_s$-sector masses exceed their reference values by roughly 145 $\%$, while the $B_s$-sector errors are about 59 $\%$. The strangeonium entries are even less realistic. Their large deviations should not be interpreted as ordinary fit residuals within the intended heavy-quark domain. The origin is quantitative. For $R=0.592$, $0.586$, and $0.676$ fm, the strange-quark mode gives $p_{\mathrm{char}}/m_s\simeq7.2$, $7.2$, and $6.3$, respectively. The expansion of the kinetic energy is therefore invalid. The explicit $1/(m_1m_2)$ factor also enhances the contact interaction, producing splittings of $326.4$ MeV in the $s\bar c$ row and approximately $3$ GeV in the $s\bar s$ row. The latter is incompatible with treating the spin interaction as a small first-order perturbation. The infrared cutoff $Q\geq2$ GeV in the coupling prescription prevents evaluation of the one-loop coupling too close to its singular region. It does not restore nonrelativistic kinematics for the strange constituent. Replacing the strange mass with a larger effective constituent mass could change the numbers substantially, but that would be a different parametrization requiring a consistent refit. Moreover, the $\eta'$ cannot be used as a pure $s\bar s$ pseudoscalar partner of the $\phi$: flavor mixing and singlet dynamics lie outside this two-body treatment. The corresponding $61.7$ MeV mass difference is therefore not a clean strangeonium hyperfine benchmark.

\section{Conclusion} \label{con}

We developed a Galilean covariant extension of the MIT bag model to describe mesons composed of heavy quark–antiquark pairs. The analysis is motivated by the relevance of Galilean symmetry in low-energy physics and builds on earlier formulations of Galilean covariance in a 5-dimensional framework, where non-relativistic dynamics can be treated in a covariant way.  A formulation of a bag-like model Lagrangian under Galilean covariance is presented. Inside the bag, quarks satisfy a Dirac-like equation, while boundary conditions ensure confinement by enforcing zero current flow at the surface. Analytical solutions for the spinor wavefunctions are obtained, along with quantization conditions for the eigenfrequencies.

The bag constant $B$ is derived from energy-momentum conservation and is interpreted as the pressure at the surface of the bag. The total energy of the system is expressed as a function of the bag radius $R$, quark mass $m_Q$, and eigenfrequencies. The model is applied to heavy quarkonium systems (e.g., charmonium and bottomonium) and is extended to open heavy mesons. Additional physical effects are incorporated through: a) minimal SU(3) gauge coupling, introducing gluon interactions; b) Hyperfine splitting, derived from spin–spin interactions; c) Color-Coulomb corrections, approximated via one-gluon exchange.

The model reproduces the spin-averaged masses reasonably well for the fitted heavy-quarkonium states, while the hyperfine splitting is well described for charmonium but remains underestimated for bottomonium. For open heavy mesons, the comparison with available reference values indicates larger deviations, particularly for systems containing strange quarks. These results indicate that the present non-relativistic Galilean-covariant formulation provides a useful phenomenological description of heavy-quark systems, while also showing limitations as relativistic effects become increasingly important.

Overall, the Galilean-covariant bag model provides a consistent framework for investigating confinement and the spectroscopy of heavy mesons within a non-relativistic setting. The limitations observed for strange and other lighter-quark systems also indicate possible directions for further improvements, including relativistic corrections and a more systematic treatment of the effective quark--gluon interaction.

\begin{acknowledgments}
 MVTM acknowledges the support of the project INCT-FNA (process 408419/2024-5) and CNPq (grant number 303075/2022-8), Brazil. 
 We thank Alisson Matheus Araujo Silva for initial discussions.
\end{acknowledgments}

\bibliographystyle{apsrev4-1}
\bibliography{ref}

@article{Takahashi,
  title={Towards the many-body theory with the Galilei invariance as a guide. 1},
  author={Takahashi, Yasushi},
  journal={Fortschritte der Physik},
  volume={36},
  number={1},
  pages={63--81},
  year={1988}
}

@book{santana2023symplectic,
  title={Symplectic Quantum Field Theory},
  author={Santana, Ademir E and Khanna, Faqir C and Santos, Alesandro F and Amorim, Ronni G Gomes and Ulhoa, S{\'e}rgio C and Vianna, J David M},
  year={2023},
  publisher={Cambridge Scholars Publishing}
}

@article{Santana,
  title={Galilei covariance and (4, 1)-de Sitter space},
  author={Santana, Ademir Eug{\^e}nio de and Khanna, Faqir C and Takahashi, Y},
  journal={Prog. theor. phys.},
  volume={99},
  number={3},
  pages={327--336},
  year={1998},
  publisher={Oxford University Press}
}

@article{santos2004galilean,
  title={Galilean covariant Lagrangian models},
  author={Santos, ES and de Montigny, M and Khanna, FC and Santana, AE},
  journal={J. Phys. A: Math. and Gen.},
  volume={37},
  number={41},
  pages={9771},
  year={2004},
  publisher={IOP Publishing}
}

@article{ParticleDataGroup:2024cfk,
    author = "Navas, S. and others",
    collaboration = "Particle Data Group",
    title = "{Review of particle physics}",
    doi = "10.1103/PhysRevD.110.030001",
    journal = "Phys. Rev. D",
    volume = "110",
    number = "3",
    pages = "030001",
    year = "2024"
}

@article{Akbar:2023eqm,
    author = "Akbar, Nosheen and Shafaq, Bushra and Zahra, Sarwat and Mir, Amna",
    title = "{Mass spectrum, root-mean-square radii, form factors, and charge radii of mesons}",
    eprint = "2303.07394",
    archivePrefix = "arXiv",
    primaryClass = "hep-ph",
    doi = "10.1007/s40042-025-01365-y",
    journal = "J. Korean Phys. Soc.",
    volume = "86",
    number = "11",
    pages = "1037--1047",
    year = "2025"
}

@article{Haxton:1980mc,
    author = "Haxton, W. C. and Heller, L.",
    title = "{The Heavy Quark - Anti-quark Potential in the {MIT} Bag Model}",
    reportNumber = "LA-UR-80-855",
    doi = "10.1103/PhysRevD.22.1198",
    journal = "Phys. Rev. D",
    volume = "22",
    pages = "1198",
    year = "1980"
}

@article{Hasenfratz:1980jv,
    author = "Hasenfratz, P. and Horgan, R. R. and Kuti, J. and Richard, J. M.",
    title = "{The Effects of Colored Glue in the QCD Motivated Bag of Heavy Quark - anti-Quark Systems}",
    reportNumber = "CERN-TH-2837",
    doi = "10.1016/0370-2693(80)90491-8",
    journal = "Phys. Lett. B",
    volume = "95",
    pages = "299--305",
    year = "1980"
}

@article{Gross:2022hyw,
    author = "Gross, Franz and others",
    title = "{50 Years of Quantum Chromodynamics}",
    eprint = "2212.11107",
    archivePrefix = "arXiv",
    primaryClass = "hep-ph",
    doi = "10.1140/epjc/s10052-023-11949-2",
    journal = "Eur. Phys. J. C",
    volume = "83",
    pages = "1125",
    year = "2023"
}

@article{Ahmed:2021atj,
    author = "Ahmed, J. and Manzoor, R. and Chang, L. and Raya, A. and Raya, K.",
    title = "{Heavy Quark Mesons: Mass Spectrum and Mass Relations}",
    doi = "10.1007/s00601-021-01624-1",
    journal = "Few Body Syst.",
    volume = "62",
    number = "3",
    pages = "39",
    year = "2021"
}

@article{Omugbe:2023ges,
    author = "Omugbe, E. and others",
    title = "{Non-relativistic Mass Spectra Splitting of Heavy Mesons Under the Cornell Potential Perturbed by Spin{\textendash}Spin, Spin{\textendash}Orbit and Tensor Components}",
    doi = "10.1007/s00601-023-01848-3",
    journal = "Few Body Syst.",
    volume = "64",
    number = "3",
    pages = "66",
    year = "2023"
}

@article{chodos1974,
  author  = {Chodos, A. and Jaffe, R. L. and Johnson, K. and Thorn, C. B. and Weisskopf, V. F.},
  title   = {New Extended Model of Hadrons},
  journal = {Phys. Rev. D},
  volume  = {9},
  pages   = {3471--3495},
  year    = {1974},
  doi     = {10.1103/PhysRevD.9.3471}
}

@article{johnson1975,
  author  = {Johnson, K.},
  title   = {The M.I.T. Bag Model},
  journal = {Acta Phys. Polonica B},
  volume  = {6},
  pages   = {865--892},
  year    = {1975}
}

@article{levy1967,
  author  = {Lévy-Leblond, J.-M.},
  title   = {Nonrelativistic Particles and Wave Equations},
  journal = {Commun. Math. Phys. },
  volume  = {6},
  pages   = {286--311},
  year    = {1967},
  doi     = {10.1007/BF01646020}
}

@article{niederer1977,
  author  = {Niederer, U.},
  title   = {Covariant Wave Equations, the Galilei Group, and the Magnetic Moment of the Electron},
  journal = {Annales de l'Institut Henri Poincaré A},
  volume  = {26},
  number  = {3},
  pages   = {313--321},
  year    = {1977}
}

@article{arrizabalaga2022,
  author        = {Arrizabalaga, Naiara and Mas, Albert and Sanz-Perela, Tom{\'a}s and Vega, Luis},
  title         = {Eigenvalue curves for generalized {MIT} bag models},
  journal       = {arXiv},
  eprint        = {2106.08348},
  archivePrefix = {arXiv},
  primaryClass  = {math.AP},
  year          = {2022},
  note          = {arXiv:2106.08348v3 [math.AP]}
}

@article{bruce2001,
  author        = {Bruce, S.},
  title         = {A {H}amiltonian for the description of a non-relativistic spin-1/2 free particle},
  journal       = {arXiv},
  eprint        = {quant-ph/0104051},
  archivePrefix = {arXiv},
  primaryClass  = {quant-ph},
  year          = {2001},
  note          = {arXiv:quant-ph/0104051v1}
}

@article{behrndt2024,
  author  = {Behrndt, Jussi and Frymark, Dale and Holzmann, Markus and Stelzer-Landauer, Christian},
  title   = {Mathematical Physics, Analysis and Geometry},
  journal = {Math. Phys. Anal. Geom.},
  volume  = {27},
  pages   = {12},
  year    = {2024},
  doi     = {10.1007/s11040-024-09484-x}
}

@article{Durval1985,
  author  = {Durval, C. and Burdet, G. and Kunzle, H. P. },
  title   = {Bargmann Structures ans Newton-Cartan Theory},
  journal = {Physical Review D},
  volume  = {31},
  number  = {8},
  pages   = {1841},
  year    = {1985}
}

@article{Aldrovandi2003,
  author  = { Aldrovandi, R. and Barbosa, A. L. and Cal\c{c}ada, M. and Pereira, J. G. },
  title   = {Kinematics of a Spacetime with an Infinite Cosmological Constant},
  journal = {Foundations of Physics},
  volume  = {33},
  number  = {4},
  pages   = {613-624},
  year    = {2003}
}

@article{eichten1978,
  author  = {Eichten, E. and Gottfried, K. and Kinoshita, T. and Lane, K. D. and Yan, T.-M.},
  title   = {Charmonium: The model},
  journal = {Phys. Rev. D},
  volume  = {17},
  pages   = {3090--3117},
  year    = {1978},
  doi     = {10.1103/PhysRevD.17.3090}
}

@article{luz2022quark,
  author  = {Luz, R. R. and Costa, Caroline S. R. and Petronilo, G. X. A. and Santana, A. E. and Amorim, R. G. G. and Paiva, R. A. S.},
  title   = {Quark-Antiquark Effective Potential in Symplectic Quantum Mechanics},
  journal = {Adv. in High Energy Phys.},
  volume  = {2022},
  pages   = {3409776},
  year    = {2022},
  doi     = {10.1155/2022/3409776},
  eprint  = {2110.12223},
  archivePrefix = {arXiv},
  primaryClass  = {hep-ph}
}

@article{luz2023fractional,
  author  = {Luz, R. R. and Abu-Shady, M. and Petronilo, G. X. A. and Santana, A. E. and Amorim, R. G. G.},
  title   = {Fractional Effective Quark-Antiquark Interaction in Symplectic Quantum Mechanics},
  journal = {Adv. in High Energy Phys.},
  volume  = {2023},
  pages   = {8366154},
  year    = {2023},
  doi     = {10.1155/2023/8366154}
}

@article{abushady2024fractional,
  author  = {Abu-Shady, M. and Luz, R. R. and Petronilo, G. X. A. and Santana, A. E. and Amorim, R. G. G.},
  title   = {On the Fractional Quark--Antiquark Confinement and Symplectic Quantum Mechanics},
  journal = {Int. J. Mod. Phys. A},
  volume  = {39},
  number  = {02n03},
  pages   = {2450011},
  year    = {2024},
  doi     = {10.1142/S0217751X24500118}
}

@misc{luz2025cornell,
  author        = {Luz, R. R. and Paiva, R. A. S. and Petronilo, G. X. A. and Santana, A. E. and Rocha Filho, T. M. and Amorim, R. G. G.},
  title         = {Symplectic Pauli-Schrödinger Equation for the Quark-Antiquark Interaction System},
  year          = {2025},
  eprint        = {2507.20045},
  archivePrefix = {arXiv},
  primaryClass  = {quant-ph},
  doi           = {10.48550/arXiv.2507.20045},
  note          = {Preprint; revised version available in 2026}
}

@article{petronilo2021size,
  title={Size-effect at finite temperature in a quark--antiquark effective model in phase space},
  author={Petronilo, GXA and Amorim, RGG and Ulhoa, SC and Santos, AF and Santana, AE and Khanna, Faqir C},
  journal={Int. J. Mod. Phys. A},
  volume={36},
  number={17},
  pages={2150121},
  year={2021},
  publisher={World Scientific}
}

@article{sebastian2023lienard,
  author  = {Sebastian, Jobin and Jamal, Mohammad Yousuf and Haque, Najmul},
  title   = {Li{\'e}nard--Wiechert potential of a heavy quarkonium moving in QGP medium},
  journal = {Phys. Rev. D},
  volume  = {107},
  number  = {5},
  pages   = {054040},
  year    = {2023},
  doi     = {10.1103/PhysRevD.107.054040},
  publisher = {American Physical Society}
}

@article{santana2001,
  author  = {Santana, A. E. and de Montigny, M. and Khanna, F. C. and Santos, E. S.},
  title   = {Galilean Covariance and Non-Relativistic Bhabha Equations},
  journal = {J. Phys. A: Math. and Gen.},
  volume  = {34},
  pages   = {10921--10933},
  year    = {2001},
  doi     = {10.1088/0305-4470/34/50/307}
}

@article{ulhoa2009,
  author  = {Ulhoa, S. C. and Khanna, F. C. and Santana, A. E.},
  title   = {Galilean Covariance and the Gravitational Field},
  journal = {Int. J. Mod. Phys. A},
  volume  = {24},
  pages   = {5287--5297},
  year    = {2009},
  doi     = {10.1142/S0217751X09045849},
  eprint  = {0902.2023},
  archivePrefix = {arXiv}
}

@article{ulhoa2011,
  author  = {Ulhoa, S. C. and Khanna, F. C. and Santana, A. E.},
  title   = {Teleparallel Formalism of Galilean Gravity},
  journal = {Gravitation and Cosmology},
  volume  = {17},
  pages   = {242--251},
  year    = {2011},
  doi     = {10.1134/S0202289311030091},
  eprint  = {1107.1275},
  archivePrefix = {arXiv}
}

@article{petronilo2019,
  author  = {Petronilo, G. X. A. and Ulhoa, S. C. and Santana, A. E.},
  title   = {Symplectic Field Theory of the Galilean Covariant Scalar and Spinor Representations},
  journal = {Ukrainian J. Phys.},
  volume  = {64},
  number  = {8},
  pages   = {719--729},
  year    = {2019},
  doi     = {10.15407/ujpe64.8.719}
}

@article{petronilo2020,
  author  = {Petronilo, G. X. A. and Ulhoa, S. C. and Santana, A. E. and Ara{\'u}jo, K. V. S. and Paiva, R. A. S. and Amorim, R. G. G.},
  title   = {The Landau Problem and Nonclassicality in Phase Space},
  journal = {Int. J. Mod. Phys. A},
  volume  = {35},
  pages   = {2050148},
  year    = {2020},
  doi     = {10.1142/S0217751X20501481}
}

@article{petronilo2021,
  author        = {Petronilo, G. X. A. and Amorim, R. G. G. and Ulhoa, S. C. and Santos, A. F. and Santana, A. E. and Khanna, F. C.},
  title         = {Galilean Covariance in Phase Space at Finite Temperature},
  year          = {2021},
  eprint        = {2104.11661},
  archivePrefix = {arXiv},
  primaryClass  = {hep-th}
}

@article{arrizabalaga2019,
  author  = {Arrizabalaga, N. and Le Treust, L. and Mas, A. and Raymond, N.},
  title   = {The MIT Bag Model as an Infinite Mass Limit},
  journal = {Journal de l'École Polytechnique --- Mathématiques},
  volume  = {6},
  pages   = {329--365},
  year    = {2019},
  doi     = {10.5802/jep.95}
}

@article{Zhang2021OpenHeavyBag,
  author        = {Zhang, Wen-Xuan and Xu, Hao and Jia, Duojie},
  title         = {Masses and magnetic moments of hadrons with one and two open heavy quarks: Heavy baryons and tetraquarks},
  journal       = {Phys. Rev. D},
  volume        = {104},
  number        = {11},
  pages         = {114011},
  year          = {2021},
  doi           = {10.1103/PhysRevD.104.114011},
  eprint        = {2109.07040},
  archivePrefix = {arXiv},
  primaryClass  = {hep-ph}
}

@article{Yan2023HiddenHeavyTetraquarks,
  author        = {Yan, Ting-Qi and Zhang, Wen-Xuan and Jia, Duojie},
  title         = {Mass spectra of hidden heavy-flavor tetraquarks with two and four heavy quarks},
  journal       = {Eur. Phys. J. C},
  volume        = {83},
  pages         = {810},
  year          = {2023},
  doi           = {10.1140/epjc/s10052-023-11956-3}
}

@article{Zhu2024TriplyHeavyTetraquarks,
  author        = {Zhu, Zhen-Hui and Zhang, Wen-Xuan and Jia, Duojie},
  title         = {Triply heavy tetraquark states: masses and other properties},
  journal       = {Eur. Phys. J. C},
  volume        = {84},
  pages         = {344},
  year          = {2024},
  doi           = {10.1140/epjc/s10052-024-12700-1},
  eprint        = {2312.01908},
  archivePrefix = {arXiv},
  primaryClass  = {hep-ph}
}

@article{Zhang2024HiddenHeavyPentaquarks,
  author        = {Zhang, Wen-Xuan and Liu, Chang-Le and Jia, Duojie},
  title         = {Study of the hidden-heavy pentaquarks and {$P_{cs}$} states},
  journal       = {Phys. Rev. D},
  volume        = {109},
  number        = {11},
  pages         = {114037},
  year          = {2024},
  doi           = {10.1103/PhysRevD.109.114037},
  eprint        = {2312.12770},
  archivePrefix = {arXiv},
  primaryClass  = {hep-ph}
}

@article{Bagchi2022GalileanGauge,
  author  = {Bagchi, Arjun and Basu, Rudranil and Islam, Minhajul and Kolekar, Kedar S. and Mehra, Aditya},
  title   = {Galilean gauge theories from null reductions},
  journal = {J. High Energy Phys.},
  volume  = {2022},
  number  = {4},
  pages   = {176},
  year    = {2022},
  doi     = {10.1007/JHEP04(2022)176},
  eprint  = {2201.12629},
  archivePrefix = {arXiv},
  primaryClass = {hep-th}
}

@article{Baiguera2024NRQFT,
  author  = {Baiguera, Stefano},
  title   = {Aspects of non-relativistic quantum field theories},
  journal = {Eur. Phys. J. C},
  volume  = {84},
  pages   = {268},
  year    = {2024},
  doi     = {10.1140/epjc/s10052-024-12630-y},
  eprint  = {2311.00027},
  archivePrefix = {arXiv},
  primaryClass = {hep-th}
}

@article{DeGrand:1975cf,
    author = "DeGrand, Thomas A. and Jaffe, R. L. and Johnson, K. and Kiskis, J. E.",
    title = "{Masses and Other Parameters of the Light Hadrons}",
    reportNumber = "MIT-CTP-475",
    doi = "10.1103/PhysRevD.12.2060",
    journal = "Phys. Rev. D",
    volume = "12",
    pages = "2060",
    year = "1975"
}

\appendix
\section{One-gluon exchange}
\subsection{Feynman rules in Coulomb gauge}

We work in Coulomb gauge. The spatial gluon propagator is
\begin{equation}
D_{ij}(\boldsymbol{q})
=
\frac{i}{\boldsymbol{q}^2}
\left(
\delta_{ij}
-
\frac{q_i q_j}{\boldsymbol{q}^2}
\right).
\label{eq:prop}
\end{equation}

The quark and the antiquark may have different masses, which we denote by $m_q\equiv m_1$ and $m_{\bar q}\equiv m_2$, respectively. The Pauli vertex in momentum space on a line of mass $m_f$, with $f=q,\bar q$, is
\begin{equation}
V_i^a
=
\frac{g}{2\,m_f}
(\boldsymbol{\sigma}\times \boldsymbol{q})_i
T^a .
\label{eq:vertex}
\end{equation}
For the antiquark line, the Pauli coupling has the opposite sign (charge conjugation). This sign is absorbed in the color factor $T_q^aT_{\bar q}^a=-C_F$ used below.

\subsection{One--gluon exchange amplitude}

The spin-dependent part of the quark--antiquark scattering amplitude is given by
\begin{equation}
i\mathcal{M}
=
\frac{g}{2m_q}\,\frac{g}{2m_{\bar q}}
(\boldsymbol{\sigma}_q \times \boldsymbol{q})_i
(\boldsymbol{\sigma}_{\bar q} \times \boldsymbol{q})_j
D_{ij}(\boldsymbol{q})
\left( T_q^a T_{\bar q}^a \right) .
\end{equation}

Substituting the propagator \eqref{eq:prop}, and using $(\boldsymbol{\sigma}\times\boldsymbol{q})\cdot\boldsymbol{q}=0$ to eliminate the longitudinal term, we find
\begin{equation}
i\mathcal{M}
=
\frac{g^2}{4m_qm_{\bar q}}
\frac{i}{\boldsymbol{q}^2}
(\boldsymbol{\sigma}_q \times \boldsymbol{q})\cdot
(\boldsymbol{\sigma}_{\bar q} \times \boldsymbol{q})
\left( T_q^a T_{\bar q}^a \right) .
\label{eq:amp1}
\end{equation}

Using the vector identity
\begin{equation}
(\boldsymbol{a}\times \boldsymbol{q})\cdot(\boldsymbol{b}\times \boldsymbol{q})
=
(\boldsymbol{a}\cdot\boldsymbol{b})\,\boldsymbol{q}^2
-
(\boldsymbol{a}\cdot\boldsymbol{q})(\boldsymbol{b}\cdot\boldsymbol{q}),
\end{equation}
Eq.~\eqref{eq:amp1} becomes
\begin{equation}
i\mathcal{M}
=
\frac{g^2}{4m_qm_{\bar q}}\,
i
\left[
\boldsymbol{\sigma}_q\cdot \boldsymbol{\sigma}_{\bar q}
-
\frac{
(\boldsymbol{\sigma}_q\cdot \boldsymbol{q})
(\boldsymbol{\sigma}_{\bar q}\cdot \boldsymbol{q})
}{\boldsymbol{q}^2}
\right]
\left( T_q^a T_{\bar q}^a \right) .
\label{eq:amp2}
\end{equation}

\subsection{Color factor}

For a color-singlet quark-antiquark state in QCD,
\begin{equation}
T_q^a T_{\bar q}^a = - C_F ,
\qquad
C_F = \frac{4}{3} .
\end{equation}

Thus Eq.~\eqref{eq:amp2} becomes
\begin{equation}
\mathcal{M}
=
-\,i
\frac{g^2}{4m_qm_{\bar q}}
\frac{4}{3}
\left[
\boldsymbol{\sigma}_q\cdot \boldsymbol{\sigma}_{\bar q}
-
\frac{
(\boldsymbol{\sigma}_q\cdot \boldsymbol{q})
(\boldsymbol{\sigma}_{\bar q}\cdot \boldsymbol{q})
}{\boldsymbol{q}^2}
\right].
\label{eq:amp3}
\end{equation}

\subsection{Central--tensor decomposition}

Writing
\begin{equation}
(\boldsymbol{\sigma}_q\cdot \boldsymbol{q})
(\boldsymbol{\sigma}_{\bar q}\cdot \boldsymbol{q})
=
\sigma_q^i \sigma_{\bar q}^j q_i q_j ,
\end{equation}
we decompose
\begin{equation}
q_i q_j
=
\frac{1}{3}\boldsymbol{q}^2 \delta_{ij}
+
\left(
q_i q_j
-
\frac{1}{3}\boldsymbol{q}^2 \delta_{ij}
\right).
\end{equation}

The first term is rotationally invariant and contributes to the spin--spin interaction, while the second term is traceless and generates the tensor force.

Using the isotropic part,
\begin{equation}
\frac{
(\boldsymbol{\sigma}_q\cdot \boldsymbol{q})
(\boldsymbol{\sigma}_{\bar q}\cdot \boldsymbol{q})
}{\boldsymbol{q}^2}
=
\frac{1}{3}
\boldsymbol{\sigma}_q\cdot \boldsymbol{\sigma}_{\bar q}
+ \text{tensor} .
\end{equation}

Substituting into Eq.~\eqref{eq:amp3}, the central contribution becomes
\begin{equation}
\boldsymbol{\sigma}_q\cdot \boldsymbol{\sigma}_{\bar q}
-
\frac{1}{3}
\boldsymbol{\sigma}_q\cdot \boldsymbol{\sigma}_{\bar q}
=
\frac{2}{3}
\boldsymbol{\sigma}_q\cdot \boldsymbol{\sigma}_{\bar q} .
\end{equation}

\subsection{Spin--spin potential}

Dropping the overall factor of $i$ and performing a Fourier transform to position space, we obtain
\begin{equation}
V_{SS}(\boldsymbol{r})
=
\frac{g^2}{4m_qm_{\bar q}}
\frac{4}{3}
\frac{2}{3}
\boldsymbol{\sigma}_q\cdot \boldsymbol{\sigma}_{\bar q}
\delta^{(3)}(\boldsymbol{r}) .
\end{equation}

Using $g^2 = 4\pi \alpha_s$, this becomes
\begin{equation}
V_{SS}(\boldsymbol{r})
=
\frac{8\pi \alpha_s}{9\,m_qm_{\bar q}}
\boldsymbol{\sigma}_q\cdot \boldsymbol{\sigma}_{\bar q}
\delta^{(3)}(\boldsymbol{r}) .
\end{equation}

Introducing the physical spin operators,
\begin{equation}
\boldsymbol{S} = \frac{\boldsymbol{\sigma}}{2},
\qquad
\boldsymbol{\sigma}_q\cdot \boldsymbol{\sigma}_{\bar q}
=
4\, \boldsymbol{S}_q\cdot \boldsymbol{S}_{\bar q},
\end{equation}
we arrive at the final result:
\begin{equation}
\boxed{
V_{SS}(\boldsymbol{r})
=
\frac{32\pi \alpha_s}{9\,m_qm_{\bar q}}
\,
\boldsymbol{S}_q\cdot \boldsymbol{S}_{\bar q}
\,
\delta^{(3)}(\boldsymbol{r})
} .
\end{equation}
For equal masses ($c\bar c$, $b\bar b$) this reduces to the $1/m^2$ form. This is the interaction $H_{SS}$ of Eq.~\eqref{eq:HSS} with $m_1=m_q$ and $m_2=m_{\bar q}$.

\section{Embedding of the Euclidean Space into the De Sitter space}

Consider $\mathcal{G}$ as an analytical manifold where each point is characterized by coordinates $x^\mu$, with $\mu = 1, 2, 3,4,5$ and the metric specified by \eqref{1}.  Additionally, let’s explore $\mathcal{E}_3$, which represents the three-dimensional Euclidean manifold. Therefore,
\begin{equation}
    \mathcal{I}_1: \textbf{A}\rightarrow A=\left(\textbf{A},A_4, \frac{\textbf{A}^2}{2A_4}\right);\quad\textbf{A}\in \mathcal{E}_3, A\in \mathcal{G},
\end{equation}
has norm $A^2=\textbf{A}-2A_4A_5=0$. The embedding $\mathcal{I}_1$ establishes a correspondence of $\mathcal{E}_3$ with vectors that have a null norm of $\mathcal{G}$. An example is the vectors associated with Galilean invariance in $\mathcal{G}$.
Two equivalent embeddings, but different from $\mathcal{I}_1$, are given by
\begin{eqnarray}
    &\mathcal{I}_2: \textbf{A}\rightarrow A=\left(\textbf{A},A_4, 0\right);\\
    &\mathcal{I}_3: \textbf{A}\rightarrow A=\left(\textbf{A},0, 0\right).
\end{eqnarray}
The norm is non-null, $A^2=\textbf{A}^2$; that is, the Euclidean vector norm.
\end{document}